\documentclass{article}
\usepackage{ijcai17}

\usepackage{times}

\usepackage{graphicx}
\usepackage{subcaption}
\graphicspath{{img/}}
\usepackage{placeins}
\usepackage{bm}

\usepackage{epsfig}
\usepackage{amssymb}
\usepackage{amsmath}
\usepackage{amsfonts}
\usepackage{color-edits}

\addauthor{xh}{blue}
\addauthor{nk}{red}

\usepackage{color}

\newcommand{\vct}[1]{\boldsymbol{#1}} % vector

\usepackage{algorithmic}    	% use algorithmic
\usepackage[ruled]{algorithm2e}	% use algorithm2e

\title{DynGEM: Deep Embedding Method for Dynamic Graphs
}
\author{
Palash Goyal${^*}$, Nitin Kamra\thanks{Nitin Kamra and Palash Goyal contribute equally to this article.}, Xinran He, Yan Liu\\
University of Southern California\\
\{palashgo, nkamra, xinranhe, yanliu.cs\}@usc.edu
}

\begin{document}

\maketitle

\begin{abstract}
\label{abstract}

Embedding large graphs in low dimensional spaces has recently attracted significant interest due to its wide applications such as graph visualization, link prediction and node classification. Existing methods focus on computing the embedding for static graphs. However, many graphs in practical applications are dynamic and evolve constantly over time. Naively applying existing embedding algorithms to each snapshot of dynamic graphs independently usually leads to unsatisfactory performance in terms of stability, flexibility and efficiency. In this work, we present an efficient algorithm DynGEM based on recent advances in deep autoencoders for graph embeddings, to address this problem. The major advantages of DynGEM include: (1) the embedding is stable over time, (2) it can handle growing dynamic graphs, and (3) it has better running time than using static embedding methods on each snapshot of a dynamic graph. We test DynGEM on a variety of tasks including graph visualization, graph reconstruction, link prediction and anomaly detection (on both synthetic and real datasets). Experimental results demonstrate the superior stability and scalability of our approach.
\vspace{-5pt}
\end{abstract}

\section{Introduction}
\label{sec:intro}

Many important tasks in network analysis involve making predictions over nodes and/or edges in a graph, which demands effective algorithms for extracting meaningful patterns and constructing predictive features.
Among the many attempts towards this goal, graph embedding, i.e., learning low-dimensional representation for each node in the graph that accurately captures its relationship to other nodes, has recently attracted much attention. 
It has been demonstrated that graph embedding is superior to alternatives in many supervised learning tasks, such as node classification, link prediction and graph reconstruction \cite{Ahmed2013,Perozzi2014,Cao2015,Tang2015,Grover2016,Ou2016}.

Various approaches have been proposed for static graph embedding~\cite{goyal2017graph}. 
Examples include SVD based models~\cite{Belkin2001,Roweis2000,Tenenbaum2000,Cao2015,Ou2016}, which decompose the Laplacian or high-order adjacency matrix to produce node embeddings.
Others include Random-walk based models~\cite{Grover2016,Perozzi2014} which create embeddings from localized random walks and many others~\cite{Tang2015,Ahmed2013,cao2016deep,niepert2016learning}.
Recently, Wang et al. designed an innovative model, SDNE, which utilizes a deep autoencoder to handle non-linearity to generate more accurate embeddings~\cite{Wang2016}.
Many other methods which handle attributed graphs and generate a unified embedding have also been proposed in the recent past~\cite{chang2015heterogeneous,huang2017accelerated,huang2017label}.

However, in practical applications, many graphs, such as social networks, are dynamic and evolve over time. For example, new links are formed (when people make new friends) and old links can disappear. 
Moreover, new nodes can be introduced into the graph (e.g., users can join the social network) and create new links to existing nodes.  Usually, we represent the dynamic graphs as a collection of snapshots of the graph at different time steps~\cite{Leskovec2007}. 

Existing works which focus on dynamic embeddings often apply static embedding algorithms to each snapshot of the dynamic graph and then rotationally align the resulting static embeddings across time steps~\cite{Hamilton2016,Kulkarni2015}.
Naively applying existing static embedding algorithms independently to each snapshot leads to unsatisfactory performance due to the following challenges:
\begin{itemize}
	\item \textbf{Stability}: The embedding generated by static methods is not stable, i.e., the embedding of graphs at consecutive time steps can differ substantially even though the graphs do not change much.
	\item \textbf{Growing Graphs}: New nodes can be introduced into the graph and create new links to existing nodes as the dynamic graph grows in time. All existing approaches assume a fixed number of nodes in learning graph embeddings and thus cannot handle growing graphs.
	\item \textbf{Scalability}: Learning embeddings independently for each snapshot leads to running time linear in the number of snapshots. As learning a single embedding is already computationally expensive, the naive approach does not scale to dynamic networks with many snapshots.
\end{itemize}
Other approaches have attempted to learn embedding of dynamic graphs by explicitly imposing a temporal regularizer to ensure temporal smoothness over embeddings of consecutive snapshots~\cite{Zhu2016}. This approach fails for dynamic graphs where consecutive time steps can differ significantly, and hence cannot be used for applications like anomaly detection. 
Moreover, their approach is a Graph Factorization (abbreviated as GF hereafter)~\cite{Ahmed2013} based model, and DynGEM outperforms these models as shown by our experiments in section \ref{sec:results}.
\cite{dai2017deep} learn embedding of dynamic graphs, although they focus on a bipartite graphs specifically for user-item interactions.

In this paper, we develop an efficient graph embedding algorithm, referred to as DynGEM, to generate stable embeddings of dynamic graphs. DynGEM employs a deep autoencoder at its core and leverages the recent advances in deep learning to generate highly non-linear embeddings. 
Instead of learning the embedding of each snapshot from scratch, DynGEM incrementally builds the embedding of snapshot at time $t$ from the embedding of snapshot at time $t-1$. Specifically, we initialize the embedding from previous time step, and then carry out gradient training. This approach not only ensures stability of embeddings across time, but also leads to efficient training as all embeddings after the first time step require very few iterations to converge.
To handle dynamic graphs with growing number of nodes, we incrementally grow the size of our neural network with our heuristic, PropSize, to dynamically determine the number of hidden units required for each snapshot.
In addition to the proposed model, we also introduce rigorous stability metrics for dynamic graph embeddings. 

On both synthetic and real-world datasets, experiment results demonstrate that our approach achieves similar or better accuracy in graph reconstruction and link prediction more efficiently than existing static approaches.
DynGEM is also applicable for dynamic graph visualization and anomaly detection, which are not feasible for many previous static embedding approaches.

\section{Definitions and Preliminaries}
\label{sec:defn}
We denote a weighted graph as $G(V, E)$ where $V$ is the vertex set and $E$ is the edge set. The weighted adjacency matrix of $G$ is denoted by $S$. If $(u,v)\in E$, we have $s_{ij} > 0$ denoting the weight of edge $(u,v)$; otherwise we have $s_{ij}=0$. We use $\vct{s_i} = [s_{i,1}, \cdots, s_{i, |V|}]$ to denote the $i$-th row of the adjacency matrix.

Given a graph $G=(V,E)$, a \emph{graph embedding} is a mapping $f:V \rightarrow \mathbb{R}^d$, namely $\vct{y_v}=f(v) \ \forall v \in V$. We require that $d \ll |V|$ and the function $f$ preserves some proximity measure defined on the graph $G$. Intuitively, if two nodes $u$ and $v$ are ``similar'' in graph $G$, their embedding $\vct{y_u}$ and $\vct{y_v}$ should be close to each other in the embedding space. We use the notation $f(G)\in \mathbb{R}^{|V|\times d}$ for the embedding matrix of all nodes in the graph $G$.

In this paper, we consider the problem of \emph{dynamic graph embedding}. We represent a dynamic graph $\mathcal{G}$ as a series of snapshots, i.e. $\mathcal{G} = \{G_1, \cdots, G_T\}$, where $G_t=(V_t, E_t)$ and $T$ is the number of snapshots. We consider the setting with growing graphs i.e. $V_t \subseteq V_{t+1}$, namely new nodes can join the dynamic graph and create links to existing nodes. We consider the deleted nodes as part of the graph with zero weights to the rest of the nodes. We assume no relationship between $E_t$ and $E_{t+1}$ and new edges can form between snapshots while existing edges can disappear.

A \emph{dynamic graph embedding} extends the concept of embedding to dynamic graphs. Given a dynamic graph $\mathcal{G}=\{G_1, \cdots, G_T\}$, a dynamic graph embedding is a time-series of mappings $\mathcal{F} = \{f_1, \cdots, f_T\}$ such that mapping $f_t$ is a graph embedding for $G_t$ and all mappings preserve the proximity measure for their respective graphs.

A successful dynamic graph embedding algorithm should create \emph{stable} embeddings over time. Intuitively, a stable dynamic embedding is one in which consecutive embeddings differ only by small amounts if the underlying graphs change a little i.e. if $G_{t+1}$ does not differ from $G_t$ a lot, the embedding outputs $Y_{t+1} = f_{t+1}(G_{t+1})$ and $Y_t = f_t(G_t)$ also change only by a small amount.

To be more specific, let $S_t(\tilde{V})$ be the weighted adjacency matrix of the induced subgraph of node set $\tilde{V}\subseteq V_t$ and $F_t(\tilde{V})\in \mathbb{R}^{|\tilde{V}|\times d}$ be the embedding of all nodes in $\tilde{V}\subseteq V_t$ of snapshot $t$.  We define the \emph{absolute stability} as
$$
\mathcal{S}_{abs}(\mathcal{F}; t) = \frac{\| F_{t+1}(V_t) - F_t(V_t) \|_F}{\| S_{t+1}(V_t) - S_t(V_t) \|_F}.
$$
In other words, the absolute stability of any embedding $\mathcal{F}$ is the ratio of the difference between embeddings to that of the difference between adjacency matrices. Since this definition of stability depends on the sizes of the matrices involved, we define another measure called \emph{relative stability} which is invariant to the size of adjacency and embedding matrices:
\begin{align*}
\mathcal{S}_{rel}(\mathcal{F}; t) = \frac{\| F_{t+1}(V_t) - F_t(V_t) \|_F}{\|F_t(V_t)\|_F} \bigg/ \frac{\| S_{t+1}(V_t) - S_t(V_t) \|_F}{\|S_t(V_t)\|_F}.
\end{align*}
We further define the \emph{stability constant}:
$$K_\mathcal{S}(\mathcal{F})=\max_{\tau, \tau'} |\mathcal{S}_{rel}(F;\tau) - \mathcal{S}_{rel}(F;\tau')|.$$
We say that a dynamic embedding $\mathcal{F}$ is stable as long as it has a small stability constant. Clearly, the smaller the $K_\mathcal{S}(\mathcal{F})$ is, the more stable the embedding $\mathcal{F}$ is. In the experiments, we use the stability constant as the metric to compare the stability of our DynGEM algorithm to other baselines.

\section{DynGEM: Dynamic Graph Embedding Model}
\label{sec:dyngem}

\begin{figure}
    \centering
    \includegraphics[width=0.5\textwidth]{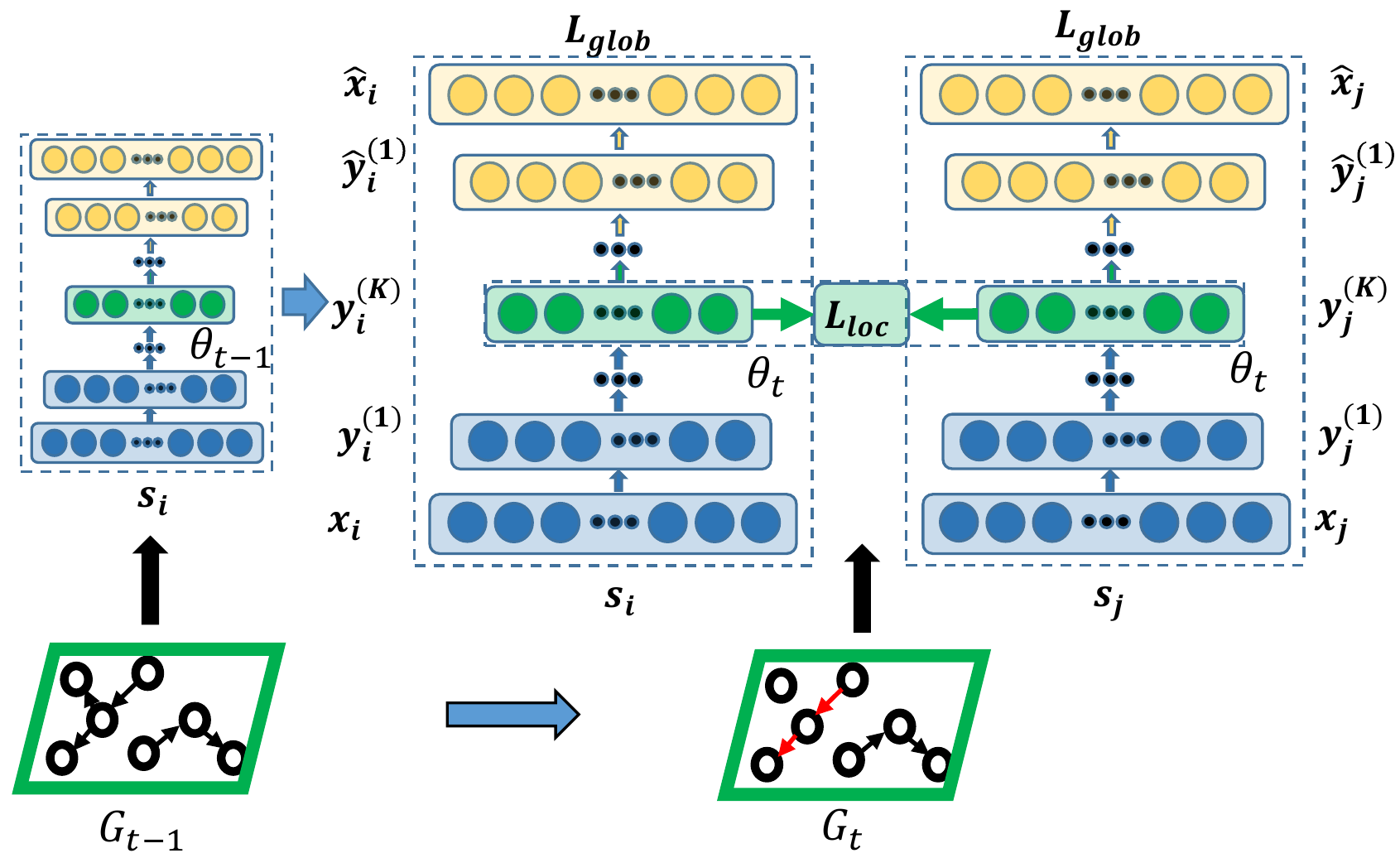}
    \caption{DynGEM: Dynamic Graph Embedding Model. The figure shows two snapshots of a dynamic graph and the corresponding deep autoencoder at each step.}
    \label{fig:dyngem}
\end{figure}

Recent advances in deep unsupervised learning have shown that autoencoders can successfully learn very complex low-dimensional representations of data for various tasks~\cite{Bengio2013}. DynGEM uses a deep autoencoder to map the input data to a highly nonlinear latent space to capture the connectivity trends in a graph snapshot at any time step.
The model is semi-supervised and minimizes a combination of two objective functions corresponding to the first-order proximity and second-order proximity respectively.
The autoencoder model is shown in Figure \ref{fig:dyngem}, and the terminology used is in Table~\ref{tab:conventions} (as in \cite{Wang2016}). The symbols with hat on top are for the decoder.

\begin{table}
	\centering
	\begin{tabular}{|c|c|}
	\hline
	Symbol & Definition \\
	\hline
	$n$ & number of vertices \\
	$K$ & number of layers \\
	$S$ = $\{\vct{s_1}, \cdots, \vct{s_n}\}$ & adjacency matrix of $G$ \\
	$X = \{\vct{x_i}\}_{i \in [n]}$ & input data \\
	$\hat{X} = \{\vct{\hat{x}_i}\}_{i \in [n]}$ & reconstructed data \\
	$Y^{(k)} = \{\vct{y_i^{(k)}}\}_{i \in [n]}$ & hidden layers \\
	$Y = Y^{(K)}$ & embedding \\
	$\vct{\theta} = \{W^{(k)}, \hat{W}^{(k)}, \vct{b^{(k)}}, \vct{\hat{b}^{(k)}}\}$ & weights, biases \\
	\hline
	\end{tabular}
	\caption{Notations for deep autoencoder}
	\label{tab:conventions}
	\vspace{-10pt}
\end{table}

The neighborhood of a node $v_i$ is given by $\vct{s_i} \in \mathbb{R}^n$. 
For any pair of nodes $v_i$ and $v_j$ from graph $G_t$, the model takes their neighborhoods as input: $\vct{x_i} = \vct{s_i}$ and $\vct{x_j} = \vct{s_j}$, and passes them through the autoencoder to generate $d$-dimensional embeddings $\vct{y_i} = \vct{y_i^{(K)}}$ and $\vct{y_j} = \vct{y_j^{(K)}}$ at the outputs of the encoder.
The decoder reconstructs the neighborhoods $\vct{\hat{x}_i}$ and $\vct{\hat{x}_j}$, from embeddings $\vct{y_i}$ and $\vct{y_j}$.

\subsection{Handling growing graphs}
\label{subsec:nodeadd}
Handling dynamic graphs of growing sizes requires a good mechanism to expand the autoencoder model, while preserving weights from previous time steps of training. A key component is to decide how the number of hidden layers and the number of hidden units should grow as more nodes are added to the graph. 
We propose a heuristic, \emph{PropSize}, to compute new layer sizes for all layers which ensures that the sizes of consecutive layers are within a certain factor of each other.

\textbf{PropSize}: We propose this heuristic to compute the new sizes of neural network layers at each time step and insert new layers if needed.
For the encoder, layer widths are computed for each pair of consecutive layers ($l_k$ and $l_{k+1}$), starting from its input layer ($l_1 = \vct{x}$) and first hidden layer ($l_2 = \vct{y^{(1)}}$) until the following condition is satisfied for each consecutive layer pair: 
\begin{align*}
size(l_{k+1}) \geq \rho \times size(l_k),
\end{align*} 
where $0 <\rho < 1$ is a suitably chosen hyperparameter. If the condition is not satisfied for any pair $(l_k, l_{k+1})$, the layer width for $l_{k+1}$ is increased to satisfy the heuristic.
Note that the size of the embedding layer $\vct{y}=\vct{y^{(K)}}$ is always kept fixed at $d$ and never expanded. If the PropSize rule is not satisfied at the penultimate and the embedding layer of the encoder, we add more layers in between (with sizes satisfying the rule) till the rule is satisfied.
This procedure is also applied to the decoder layers starting from the output layer ($\vct{\hat{x}}$) and continuing inwards towards the embedding layer (or $\vct{\hat{y}} = \vct{\hat{y}^{(K)}}$) to compute new layer sizes.

After deciding the number of layers and the number of hidden units in each layer, we adopt \emph{Net2WiderNet} and \emph{Net2DeeperNet} approaches from \cite{Chen2015} to expand the deep autoencoder.
Net2WiderNet allows us to widen layers i.e. add more hidden units to an existing neural network layer, while approximately preserving the function being computed by that layer.
Net2DeeperNet inserts a new layer between two existing layers by making the new intermediate layer closely replicate the identity mapping. This can be done for ReLU activations but not for sigmoid activations.

The combination of widening and deepening the autoencoder with PropSize, Net2WiderNet and Net2DeeperNet at each time step, allows us to work with dynamic graphs with growing number of nodes over time and gives a remarkable performance as shown by our experiments.

\subsection{Loss function and training}
\label{subsec:lossfn}

To learn the model parameters, a weighted combination of three objectives is minimized at each time step:
\begin{align*}
	L_{net} &= L_{glob} + \alpha L_{loc} + \nu_1 L_1 + \nu_2 L_2,
\end{align*}
where $\alpha, \nu_1$ and $\nu_2$ are hyperparameters appropriately chosen as relative weights of the objective functions.
$L_{loc} = \sum_{i,j}^n s_{ij} \|\vct{y_i} - \vct{y_j}\|_2^2$ is the first-order proximity which corresponds to local structure of the graph.
$L_{glob} = \sum_{i=1}^n \|(\vct{\hat{x}_i} - \vct{x_i}) \odot \vct{b_i}\|_2^2 = \|(\hat{X} - X) \odot B\|_F^2$ is the second-order proximity which corresponds to global neighborhood of each node in the graph and is preserved by an unsupervised reconstruction of the neighborhood of each node. $\vct{b_i}$ is a vector with $b_{ij}=1$ if $s_{ij}=0$ else $b_{ij}=\beta>1$. This penalizes inaccurate reconstruction of an observed edge $e_{ij}$ more than that of unobserved edges.
Regularizers $L_1 = \sum_{k=1}^K \left( \|W^{(k)}\|_1 + \|\hat{W}^{(k)}\|_1 \right)$ \footnote{$\|W\|_1$ represents sum of absolute values of entries of $W$} and $L_2 = \sum_{k=1}^K \left( \|W^{(k)}\|_F^2 + \|\hat{W}^{(k)}\|_F^2 \right)$ are added to encourage sparsity in the network weights and to prevent the model from overfitting the graph structure respectively.
DynGEM learns the parameters $\vct{\theta_t}$ of this deep autoencoder at each time step $t$, and uses $Y_t^{(K)}$ as the embedding output for graph $G_t$.

\subsection{Stability by reusing previous step embedding}
\label{subsec:reuse}

For a dynamic graph $\mathcal{G} = \{G_1, \cdots, G_T\}$, we train the deep autoencoder model fully on $G_1$ using random initialization of parameters $\vct{\theta_1}$.
For all subsequent time steps, we initialize our model with parameters $\vct{\theta_t}$ from the previous time step parameters $\vct{\theta_{t-1}}$, before widening/deepening the model.
This results in direct knowledge transfer of structure from $f_{t-1}$ to $f_t$, so the model only needs to learn about changes between $G_{t-1}$ and $G_t$.
Hence the training converges very fast in a few iterations for time steps: $\{2, \cdots, T\}$.
More importantly, it guarantees stability by ensuring that embedding $Y_t$ stays close to $Y_{t-1}$.
Note that unlike \cite{Zhu2016} we do not impose explicit regularizers to keep the embeddings at time steps $t-1$ and $t$ close. Since if the graph snapshots at times $t-1$ and $t$ differ significantly, then so should the corresponding embeddings $f_{t-1}$ and $f_t$.
Our results in section \ref{sec:results} show the superior stability and faster runtime of our method over other baselines.

\subsection{Techniques for scalability}
\label{subsec:scale}

Previous deep autoencoder models~\cite{Wang2016} for static graph embeddings use sigmoid activation function and are trained with stochastic gradient descent (SGD).

We use ReLU in all autoencoder layers to support weighted graphs since ReLU can construct arbitrary positive entries of $\vct{s_i}$. It also accelerates training, since the derivative of ReLU is straightforward to compute, whereas the derivative of sigmoid requires computing exponentials~\cite{Glorot2011}.
Lastly, ReLU allows gradients from both objectives $L_{loc}$ and $L_{glob}$ to propagate effectively through the encoder and averts the vanishing gradient effect~\cite{Glorot2011} which we observe for sigmoid.

We also use nesterov momentum~\cite{Sutskever2013} with properly tuned hyperparameters, which converges much faster as opposed to using pure SGD.
Lastly, we observed better performance on all tasks with a combination of L1-norm and L2-norm regularizers.

The pseducode of learning DynGEM model for a single snapshot of the dynamic graph is shown in algorithm \ref{alg:dyngem}. The pseduocode can be called repeatedly on each snapshot in the dynamic graph to generate the dynamic embedding.

\begin{algorithm}[tb]
    \caption{Algorithm: DynGEM}
    \label{alg:dyngem}
\begin{algorithmic}[1]
    \STATE {\bfseries Input:} Graphs $G_t=(V_t,E_t), G_{t-1}=(V_{t-1},E_{t-1})$, Embedding $Y_{t-1}$, autoencoder weights $\bm{\theta}_{t-1}$
    \STATE {\bfseries Output:} Embedding $Y_t$
    
    \STATE From $G_t$, generate adjacency matrix $S$, penalty matrix $B$
    \STATE Create the autoencoder model with initial architecture
    \STATE Initialize $\bm{\theta}_t$ randomly \textbf{if} $t = 1$, \textbf{else} $\bm{\theta_t} = \bm{\theta}_{t-1}$
    \IF {$|V_t| > |V_{t-1}|$}
        \STATE Compute new layer sizes with PropSize heuristic
        \STATE Expand autoencoder layers and/or insert new layers
    \ENDIF
    \STATE Create set $\mathcal{S} = \{(s_i, s_j)\}$ for each $e=(v_i,v_j) \in E_t$
    \FOR {$i = 1, 2, \ldots$}
        \STATE Sample a minibatch $M$ from $\mathcal{S}$
        \STATE Compute gradients $\nabla_{\bm{\theta}_t}L_{net}$ of objective $L_{net}$ on $M$
        \STATE Do gradient update on $\bm{\theta}_t$ with nesterov momentum
    \ENDFOR
\end{algorithmic}
\end{algorithm}

\section{Experiments}
\label{sec:exp}
\subsection{Datasets}
\label{subsec:data}
We evaluate the performance of our DynGEM on both synthetic and  real-world dynamic graphs. The datasets are summarized in Table \ref{tab:dataStats}.
\begin{table}
	\centering
	\begin{tabular}{|c|c|c|c|}
	\hline
 	& $|V|$ & $|E|$ & $T$\\
	\hline
	SYN & 1,000 & 79,800-79,910 & 40\\
	HEP-TH & 1,424-7,980 & 2,556-21,036 & 60\\
	AS & 7716 & 10,695-26,467& 100\\
	ENRON & 184 & 63-591 & 128\\
	\hline
	\end{tabular}
	\caption{Statistics of datasets. $|V|$, $|E|$ and $T$ denote the number of nodes, edges and length of time series respectively.}
	\label{tab:dataStats}
	\vspace{-10pt}
\end{table}

\textbf{Synthetic Data (SYN)}: We generate synthetic dynamic graphs using Stochastic Block Model \cite{Yuchung1987}.
The first snapshot of the dynamic graph is generated to have three equal-sized communities with in-block probability 0.2 and cross-block probability 0.01. To generate subsequent graphs, we randomly pick nodes at each time step and move them to another community.
We use SYN to visualize the changes in embeddings as nodes change communities.

\textbf{HEP-TH} \cite{Gehrke2003}: The original dataset contains abstracts of paper in High Energy Physics Theory conference in the period from January 1993 to April 2003.
For each month, we create a collaboration network using all papers published upto that month.
We take the first five years data and generate a time series containing 60 graphs with number of nodes increasing from $1,424$ to $7,980$.

\textbf{Autonomous Systems(AS)} \cite{snapnets}: This is a communication network of who-talks-to-whom from the BGP (Border Gateway Protocol) logs.
The dataset contains 733 instances spanning from November 8, 1997 to January 2, 2000.
For our evaluation, we consider a subset of this dataset which contains the first 100 snapshots.

\textbf{ENRON} \cite{Enron2004}: The dataset contains emails between employees in Enron Inc. from Jan 1999 to July 2002.
We process the graph as done in \cite{Park2009} by considering email communication only between top executives for each week starting from Jan 1999.

\subsection{Algorithms and Evaluation Metrics}
\label{subsec:baselines}
We compare the performance of the following dynamic embedding algorithms on several tasks:
\begin{itemize}
	\item SDNE~\footnote{We replaced the sigmoid activation in all our SDNE baselines with ReLU activations for scalability and faster training times.}: We apply SDNE independently to each snapshot of the dynamic network.
	\item {[SDNE/GF]}$_{align}$: We first apply SDNE or GF algorithm independently to each snapshot and rotate the embedding as in~\cite{Hamilton2016} for alignment.
	\item GF$_{init}$: We apply GF algorithm whose embedding at time $t$ is initialized from the embedding at time $t-1$.
	\item DynGEM: Our algorithm for dynamic graph embedding. We set the embedding dimension $d=20$ for ENRON and $100$ for all other datasets. We use two hidden layers in the deep autoencoder with initial sizes (later they could expand) for each dataset as: ENRON = $[100, 80]$, \{HEP-TH, AS, SYN\} = $[500, 300]$. The neural network structures are chosen by an informal search over a set of architectures. We set $\rho = 0.3$, step-size decay for SGD = $10^{-5}$, Momentum coefficient = $0.99$. The other parameters are set via grid search with appropriate cross validation as follows: $\alpha\in[10^{-6}, 10^{-5}]$, $\beta\in[2,5]$ and $\nu_1\in[10^{-4}, 10^{-6}]$ and $\nu_2\in[10^{-3}, 10^{-6}]$.
\end{itemize}

In our experiments, we evaluate the performance of above models on tasks of graph reconstruction, link prediction, embedding stability and anomaly detection. For the first two tasks, graph reconstruction and link prediction, we use Mean Average Precision (MAP) as our metric (see \cite{Wang2016} for definition). To evaluate the stability of the dynamic embedding, we use the stability constant $K_\mathcal{S}(\mathcal{F})$ defined in section~\ref{sec:defn}. All experiments are performed on a Ubuntu 14.04.4 LTS system with 32 cores, 128 GB RAM and a clock speed of 2.6 GHz. The GPU used is Nvidia Tesla K40C.

\section{Results and Analysis}
\label{sec:results}

\subsection{Graph Reconstruction}
\label{subsec:reconst}
Embeddings as a good low-dimensional representations of a graph are expected to accurately reconstruct the graph. 
We reconstruct the graph edges between pairs of vertices from the embeddings, using the decoder from our autoencoder model. 
We rank the pairs of vertices according to their corresponding reconstructed proximity. 
Then, we calculate the ratio of real links in top $k$ pairs of vertices as the reconstruction precision.

\begin{table}[!htbp]
	\centering
	\begin{tabular}{|c|c|c|c|c|}
	\hline
 	& SYN & HEP-TH & AS & ENRON\\
	\hline
	GF$_{align}$ & 0.119 & 0.49 & 0.164 & 0.223\\
	GF$_{init}$ & 0.126 & \textbf{0.52} & 0.164 & 0.31\\
	SDNE$_{align}$ & 0.124 & 0.04 & 0.07 & 0.141\\
	SDNE & \textbf{0.987} & 0.51 & 0.214 & 0.38\\
	DynGEM & \textbf{0.987} & 0.491 & \textbf{0.216} & \textbf{0.424}\\
	\hline
	\end{tabular}
	\caption{Average MAP of graph reconstruction.}
	\label{tab:reconsMAP}
	\vspace{-5pt}
\end{table}

The graph reconstruction MAP metric averaged over snapshots on our datasets are shown in Table \ref{tab:reconsMAP}. The results show that DynGEM outperforms all Graph Factorization based baselines except on HEP-TH where its performance is comparable with the baselines.

\subsection{Link Prediction}
\label{subsec:predict}
Another important application of graph embedding is link prediction which tests how well a model can predict unobserved edges. 
A good representation of the network should not only be able to reconstruct the edges visible to it during training but should also be able to predict edges which are likely but missing in the training data.
\begin{table}[!htbp]
	\centering
	\begin{tabular}{|c|c|c|c|c|}
	\hline
 	& SYN & HEP-TH & AS & ENRON\\
	\hline
	GF$_{align}$ & 0.027 & 0.04 & 0.09 & 0.021\\
	GF$_{init}$ & 0.024 & 0.042 & 0.08 & 0.017\\
	SDNE$_{align}$ & 0.031 & 0.17 & 0.1 & 0.06\\
	SDNE & 0.034 & 0.1 & 0.09 & 0.081\\
	DynGEM & \textbf{0.194} & \textbf{0.26} & \textbf{0.21} & \textbf{0.084}\\
	\hline
	\end{tabular}
	\caption{Average MAP of link prediction.}
	\label{tab:lpredMAP}
	\vspace{-5pt}
\end{table}
To test this, we randomly hide 15\% of the network edges at time $t$ (call it $G_{t_{\text{hidden}}}$).
We train a dynamic embedding using snapshots $\{G_1, \cdots, G_{t-1}, G\backslash G_{t_{\text{hidden}}}\}$ and predict the hidden edges at snapshot $t$. We predict the most likely (highest weighted) edges which are not in the observed set of edges as the hidden edges.
The predictions are then compared against $G_{t_{\text{hidden}}}$ to obtain the precision scores. The prediction accuracy averaged over $t$ from $1$ to $T$ is shown in Table \ref{tab:lpredMAP}. We observe that DynGEM is able to predict missing edges better than the baselines on all datasets.
Since Graph Factorization based approaches perform consistently worse than SDNE based approach and our DynGEM algorithm, we only present results comparing our DynGEM to SDNE based algorithms for the remaining tasks due to space constraints.

\subsection{Stability of Embedding Methods}
\label{subsec:stability}

\begin{table}
	\centering
	\begin{tabular}{|c|c|c|c|c|}
	\hline
 	& SYN & HEP-TH & AS & ENRON\\
	\hline
	SDNE & 0.18 & 14.715 & 6.25 & 19.722\\
	SDNE$_{align}$ & 0.11 & 8.516 & 2.269 & 18.941\\
	DynGEM & \textbf{0.008} & \textbf{1.469} & \textbf{0.125} & \textbf{1.279}\\
	\hline
	\end{tabular}
	\caption{Stability constants $K_{\mathcal{S}}(F)$ of embeddings on dynamic graphs.}
	\label{tab:stab}
	\vspace{-10pt}
\end{table}

Stability of the embeddings is crucial for tasks like anomaly detection.
We evaluate the stability of our model and compare to other methods on four datasets in terms of stability constants in Table \ref{tab:stab}.
Our model substantially outperforms other models and provides stable embeddings along with better graph reconstruction performance (see table \ref{tab:reconsMAP} in section \ref{subsec:reconst} for reconstruction errors at this stability). In the next section, we show that we can utilize this stability for visualization and detecting anomalies in real dynamic networks.

\subsection{Visualization}
\label{subsec:viz}

\begin{figure}[t!]
    \centering
    \begin{subfigure}[thb]{0.23\textwidth}
        \centering
		\includegraphics[width=1.0\textwidth]{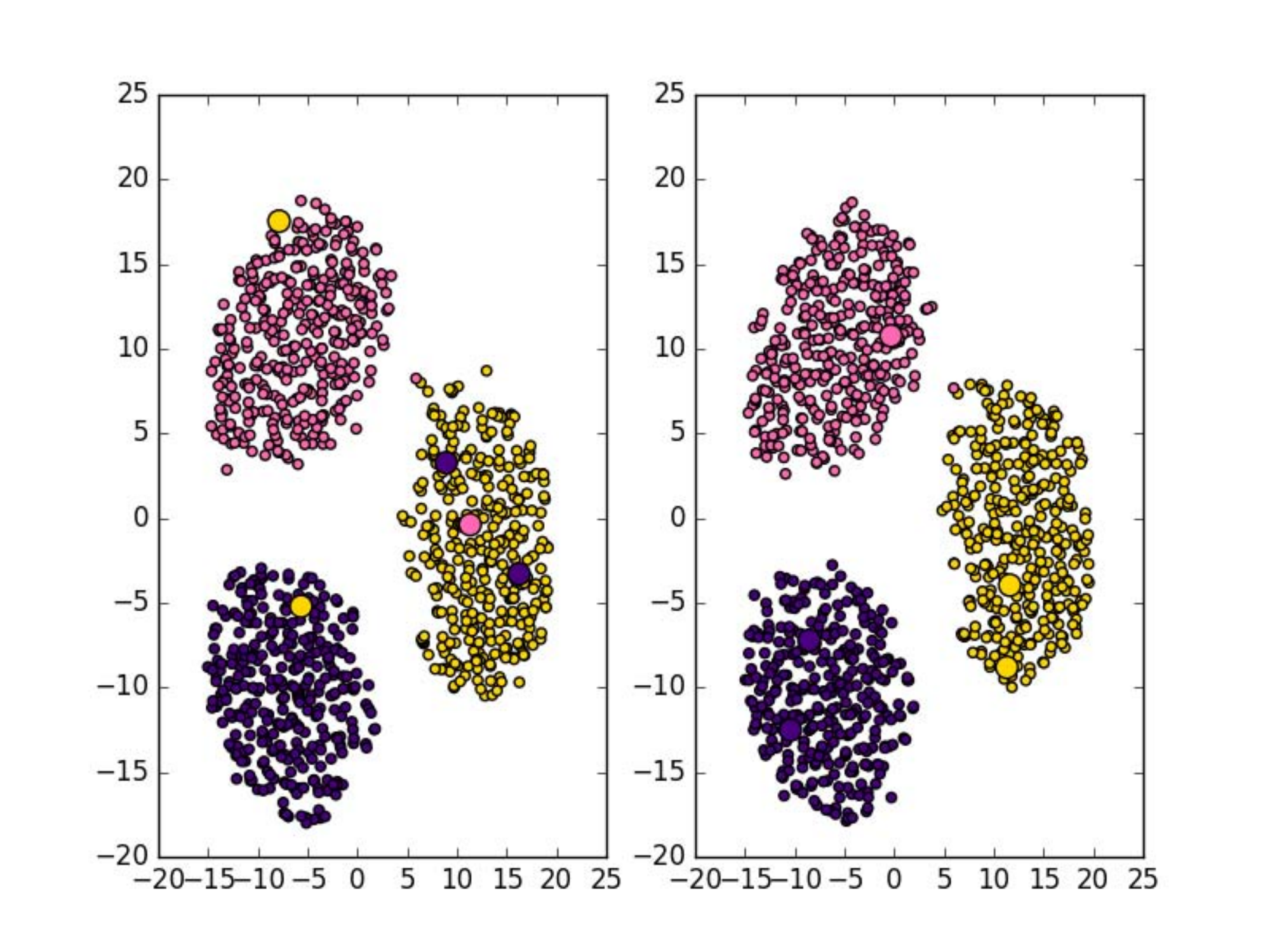}
		\caption{DynGEM time step with 5 nodes jumping out of 1000}
        \label{fig:viz_DynGEM_1000,5,2}
    \end{subfigure}
    ~
    \begin{subfigure}[thb]{0.23\textwidth}
        \centering
		\includegraphics[width=1.0\textwidth]{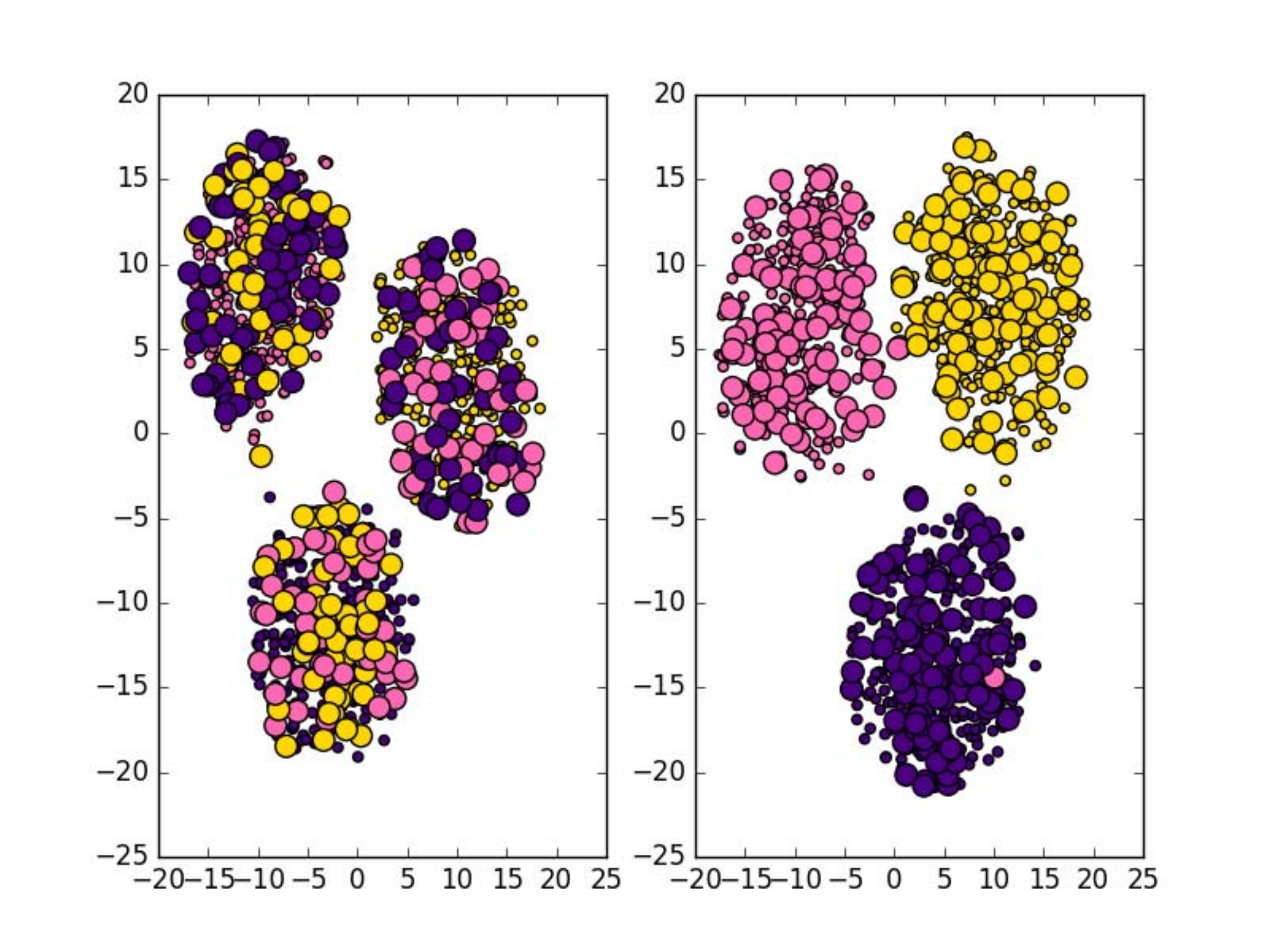}
		\caption{DynGEM time step with 300 nodes jumping out of 1000}
        \label{fig:viz_DynGEM_1000,300,2}
    \end{subfigure}
    \caption{2D visualization of $100$-dimensional embeddings for SYN dataset, when nodes change their communities over a time step. A point in any plot represents the embedding of a node in the graph, with the color indicating the node community. Small (big) points are nodes which didn't (did) change communities. Each big point is colored according to its final community color.}
    \label{fig:viz}
\end{figure}

One important application of graph embedding is graph visualization. We carry out experiments on SYN dataset with known community structure.
We apply t-SNE~\cite{Maaten2008} to the embedding generated by DynGEM at each time step to plot the resulting 2D embeddings. To avoid instability of visualization over time steps, we initialize t-SNE with identical random state for all time steps.

Figure \ref{fig:viz} illustrates the results for 2D visualization of $100$-dimensional embeddings for SYN dataset, when nodes change their communities over a single time step.
The left (right) plot in each subfigure shows the embedding before (after) the nodes change their communities.
A point in any plot represents the embedding of a node in the graph with the color indicating the node community.
Small (big) points are nodes which didn't (did) change communities. Each big point is colored according to its final community color.

We observe that the DynGEM embeddings of the nodes which changed communities, follow the changes in community structure accurately without disturbing the embeddings of other nodes, even when the fraction of such nodes is very high (see figure \ref{fig:viz_DynGEM_1000,300,2} where $30$\% nodes change communities). This strongly demonstrates the stability of our technique.

\subsection{Application to Anomaly Detection}
\label{subsec:anomaly}

\begin{figure}
    \centering
	\includegraphics[width=0.45\textwidth]{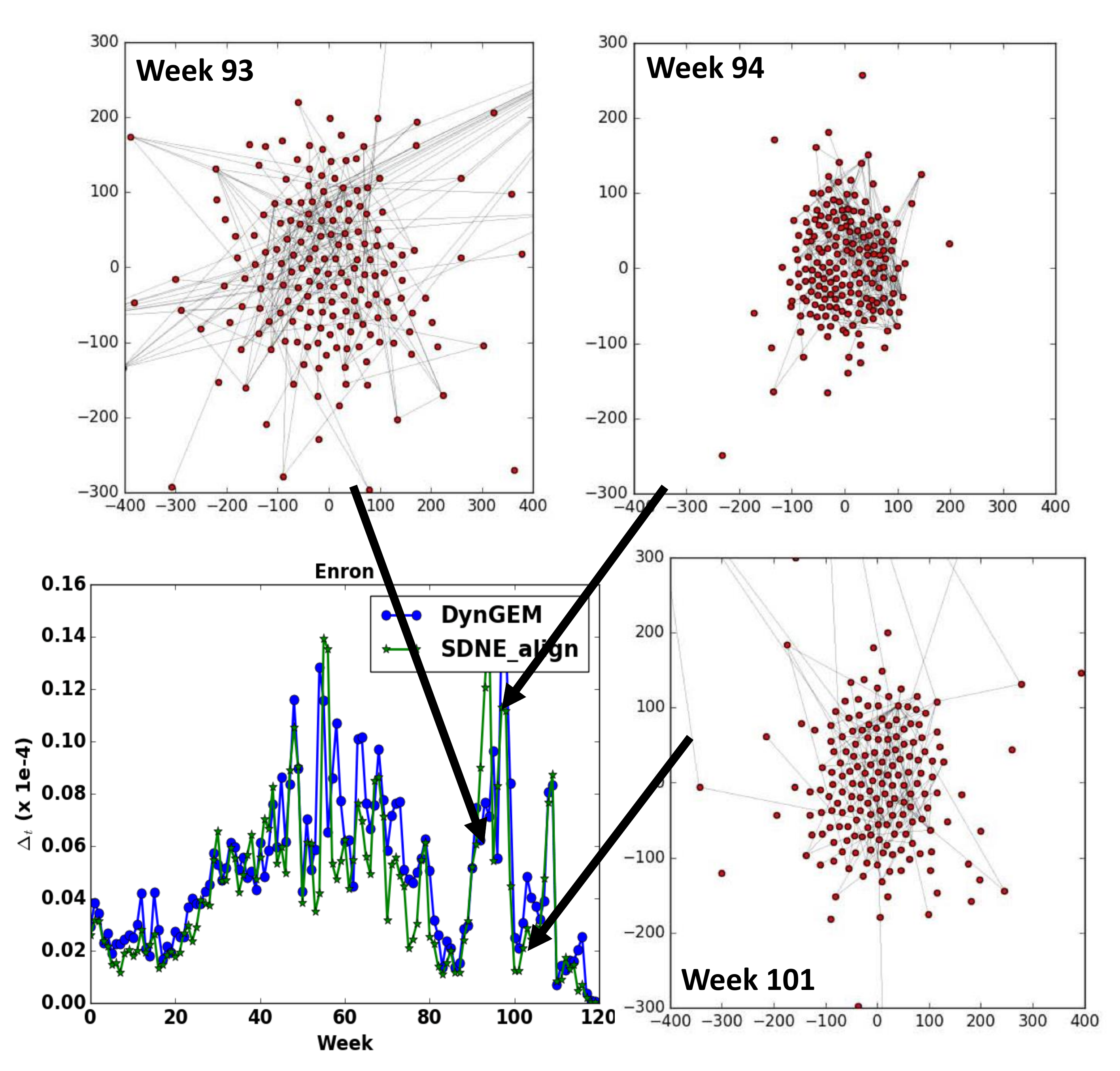}
	\caption{Results of anomaly detection on Enron and visualization of embeddings for weeks 93, 94 and 101 on Enron dataset}
   \label{fig:viz_enron}
\end{figure}

Anomaly detection is an important application for detecting malicious activity in networks.
We apply DynGEM on the Enron dataset to detect anomalies and compare our results with the publicly known events occurring to the company observed by \cite{Sun2007}.

We define $\Delta_t$ as the change in embedding between time $t$ and $t-1$: $\Delta_t = \|F_{t+1}(V_t) - F_{t}(V_t)\|_F$, and this quantity can be thresholded to detect anomalies. The plot of $\Delta_t$ with time on Enron dataset is shown in Figure \ref{fig:viz_enron}.

In the figure, we see three major spikes around week 45, 55 and 94 which correspond to Feb 2001, June 2001 and Jan 2002. These months were associated with the following events: Jeffrey Skilling took over as CEO in Feb 2001; Rove divested his stocks in energy in June 2001 and CEO resignation and crime investigation by FBI began in Jan 2002.
We also observe some peaks leading to each of these time frames which indicate the onset of these events. 
Figure \ref{fig:viz_enron} shows embedding visualizations around week 94. A spread out embedding can be observed for weeks 93 and 101, corresponding to low communication among employees. On the contrary, the volume of communication grew significantly in week 94 (shown by the highly compact embedding).

\subsection{Effect of Layer Expansion}
We evaluate the effect of layer expansion on HEP-TH data set.
For this purpose, we run our model DynGEM, with and without layer expansion.
We observe that without layer expansion, the model achieves an average MAP of 0.46 and 0.19 for graph reconstruction and link prediction respectively.
Note that this is significantly lower than the performance of DynGEM with layer expansion which obtains 0.491 and 0.26 for the respective tasks.
Also note that for SDNE and SDNE$_{align}$, we select the best model at each time step.
Using PropSize heuristic obviates this need and automatically selects a good neural network size for subsequent time steps.

\subsection{Scalability}
\label{subsec:timing}
\begin{table}
	\centering
	\begin{tabular}{|c|c|c|c|c|}
	\hline
	& SYN & HEP-TH & AS & ENRON\\
	\hline
	SDNE$_{align}$ & 56.6 min & 71.4 min & 210 min & 7.69 min\\
	DynGEM & 13.8 min &  25.4 min & 80.2 min & 3.48 min\\
	Speedup & 4.1 & 2.81 & 2.62 & 2.21\\
	Speedup$_{exp}$ & 4.8 & 3 & 3 & 3\\
	\hline
	\end{tabular}
	\caption{Computation time of embedding methods for the first forty time steps on each dataset. Speedup$_{exp}$ is the expected speedup based on model parameters.}
	\label{tab:timing}
	\vspace{-5pt}
\end{table}
\begin{table}
	\centering
	\begin{tabular}{|c|c|c|c|c|}
	\hline
	& T=5 & T=10 & T=20 & T=40\\
	\hline
	SDNE$_{align}$  & 6.99 min & 14.24 min & 26.6 min & 56.6 min\\
	DynGEM  & 2.63 min & 4.3 min & 7.21 min & 13.8 min\\
	Speedup  & 2.66 & 3.31 & 3.69 & 4.1\\
	\hline
	\end{tabular}
	\caption{Computation time of embedding methods on SYN while varying the length of time series $T$.}
	\label{tab:timing_2}
	\vspace{-10pt}
\end{table}

We now compare the time taken to learn different embedding models.
From Table \ref{tab:timing}, we observe that DynGEM is significantly faster than SDNE$_{align}$.
We do not compare it with Graph Factorization based methods because although fast, they are vastly outperformed by deep autoencoder based models.
Assuming $n_s$ iterations to learn a single snapshot embedding from scratch and $n_i$ iterations to learn embeddings when initialized with previous time step embeddings, the expected speedup for a dynamic graph of length $T$ is defined as $\frac{T n_s}{n_s + (T-1) n_i}$ (ignoring other overheads). We compare the observed speedup with the expected speedup.
In Table \ref{tab:timing_2}, we show that our model achieves speedup closer to the expected speedup as the number of graph snapshots increase due to diminished effect of overhead computations (e.g. saving, loading, expansion and initialization of the model, weights and the embedding). Our experiment results show that DynGEM achieves consistent 2-3X speed up across a variety of different networks.

\section{Conclusion}
\label{sec:conc}

In this paper, we propose DynGEM, a fast and efficient algorithm to construct stable embeddings for dynamic graphs.
It uses a dynamically expanding deep autoencoder to capture highly nonlinear first-order and second-order proximities of the graph nodes.
Moreover, our model utilizes information from previous time steps to speed up the training process by incrementally learning embeddings at each time step.
Our experiments demonstrate the stability of our technique across time and prove that our method maintains its competitiveness on all evaluation tasks e.g., graph reconstruction, link prediction and visualization.
We showed that DynGEM preserves community structures accurately, even when a large fraction of nodes ($\sim 30\%$) change communities across time steps.
We also applied our technique to successfully detect anomalies, which is a novel application of dynamic graph embedding.
DynGEM shows great potential for many other graph inference applications such as node classification, clustering etc., which we leave as future work.

There are several directions of future work. Our algorithm ensures stability by initializing from the weights learned from previous time step.
We plan to extend it to incorporate the stability metric explicitly with modifications ensuring satisfactory performance on anomaly detection.
We also hope to provide theoretical insight into the model and obtain bounds on performance.

\section*{Acknowledgments}
This work is supported in part by NSF Research Grant IIS-1254206 and IIS-1619458. The views and conclusions are those of the authors and should not be interpreted as representing the official policies
of the funding agency, or the U.S. Government.
The work was also supported in part by USC Viterbi Graduate PhD fellowship.

\newpage
\bibliographystyle{named}
\bibliography{names,conferences,bibfile}

\begin{thebibliography}{}

\bibitem[\protect\citeauthoryear{Ahmed \bgroup \em et al.\egroup
  }{2013}]{Ahmed2013}
Amr Ahmed, Nino Shervashidze, Shravan Narayanamurthy, Vanja Josifovski, and
  Alexander~J Smola.
\newblock Distributed large-scale natural graph factorization.
\newblock In {\em 22nd Intl. World Wide Web Conference}, pages 37--48, 2013.

\bibitem[\protect\citeauthoryear{Belkin and Niyogi}{2001}]{Belkin2001}
Mikhail Belkin and Partha Niyogi.
\newblock Laplacian eigenmaps and spectral techniques for embedding and
  clustering.
\newblock In {\em Proc. 13th Advances in Neural Information Processing
  Systems}, pages 585--591, 2001.

\bibitem[\protect\citeauthoryear{Bengio \bgroup \em et al.\egroup
  }{2013}]{Bengio2013}
Yoshua Bengio, Aaron Courville, and Pascal Vincent.
\newblock Representation learning: A review and new perspectives.
\newblock {\em IEEE transactions on pattern analysis and machine intelligence},
  35(8):1798--1828, 2013.

\bibitem[\protect\citeauthoryear{Cao \bgroup \em et al.\egroup
  }{2015}]{Cao2015}
Shaosheng Cao, Wei Lu, and Qiongkai Xu.
\newblock Grarep: Learning graph representations with global structural
  information.
\newblock In {\em Proc. 21st Intl. Conf. on Knowledge Discovery and Data
  Mining}, pages 891--900, 2015.

\bibitem[\protect\citeauthoryear{Cao \bgroup \em et al.\egroup
  }{2016}]{cao2016deep}
Shaosheng Cao, Wei Lu, and Qiongkai Xu.
\newblock Deep neural networks for learning graph representations.
\newblock In {\em AAAI}, pages 1145--1152, 2016.

\bibitem[\protect\citeauthoryear{Chang \bgroup \em et al.\egroup
  }{2015}]{chang2015heterogeneous}
Shiyu Chang, Wei Han, Jiliang Tang, Guo-Jun Qi, Charu~C Aggarwal, and Thomas~S
  Huang.
\newblock Heterogeneous network embedding via deep architectures.
\newblock In {\em Proceedings of the 21th ACM SIGKDD International Conference
  on Knowledge Discovery and Data Mining}, pages 119--128. ACM, 2015.

\bibitem[\protect\citeauthoryear{Chen \bgroup \em et al.\egroup
  }{2015}]{Chen2015}
Tianqi Chen, Ian Goodfellow, and Jonathon Shlens.
\newblock Net2net: Accelerating learning via knowledge transfer.
\newblock {\em arXiv preprint arXiv:1511.05641}, 2015.

\bibitem[\protect\citeauthoryear{Dai \bgroup \em et al.\egroup
  }{2017}]{dai2017deep}
Hanjun Dai, Yichen Wang, Rakshit Trivedi, and Le~Song.
\newblock Deep coevolutionary network: Embedding user and item features for
  recommendation.
\newblock 2017.

\bibitem[\protect\citeauthoryear{Gehrke \bgroup \em et al.\egroup
  }{2003}]{Gehrke2003}
Johannes Gehrke, Paul Ginsparg, and Jon Kleinberg.
\newblock Overview of the 2003 kdd cup.
\newblock {\em ACM SIGKDD Explorations Newsletter}, 5(2):149--151, 2003.

\bibitem[\protect\citeauthoryear{Glorot \bgroup \em et al.\egroup
  }{2011}]{Glorot2011}
Xavier Glorot, Antoine Bordes, and Yoshua Bengio.
\newblock Deep sparse rectifier neural networks.
\newblock In {\em Proc. 14th Intl. Conf. on Artificial Intelligence and
  Statistics}, page 275, 2011.

\bibitem[\protect\citeauthoryear{Goyal and Ferrara}{2017}]{goyal2017graph}
Palash Goyal and Emilio Ferrara.
\newblock Graph embedding techniques, applications, and performance: A survey.
\newblock {\em arXiv preprint arXiv:1705.02801}, 2017.

\bibitem[\protect\citeauthoryear{Grover and Leskovec}{2016}]{Grover2016}
Aditya Grover and Jure Leskovec.
\newblock node2vec: Scalable feature learning for networks.
\newblock In {\em Proc. 22nd Intl. Conf. on Knowledge Discovery and Data
  Mining}, pages 855--864, 2016.

\bibitem[\protect\citeauthoryear{Hamilton \bgroup \em et al.\egroup
  }{2016}]{Hamilton2016}
William~L Hamilton, Jure Leskovec, and Dan Jurafsky.
\newblock Diachronic word embeddings reveal statistical laws of semantic
  change.
\newblock {\em arXiv preprint arXiv:1605.09096}, 2016.

\bibitem[\protect\citeauthoryear{Huang \bgroup \em et al.\egroup
  }{2017a}]{huang2017accelerated}
Xiao Huang, Jundong Li, and Xia Hu.
\newblock Accelerated attributed network embedding.
\newblock In {\em Proceedings of the 2017 SIAM International Conference on Data
  Mining}, pages 633--641. SIAM, 2017.

\bibitem[\protect\citeauthoryear{Huang \bgroup \em et al.\egroup
  }{2017b}]{huang2017label}
Xiao Huang, Jundong Li, and Xia Hu.
\newblock Label informed attributed network embedding.
\newblock In {\em Proceedings of the Tenth ACM International Conference on Web
  Search and Data Mining}, pages 731--739. ACM, 2017.

\bibitem[\protect\citeauthoryear{Klimt and Yang}{2004}]{Enron2004}
Bryan Klimt and Yiming Yang.
\newblock The enron corpus: A new dataset for email classification research.
\newblock In {\em European Conference on Machine Learning}, pages 217--226,
  2004.

\bibitem[\protect\citeauthoryear{Kulkarni \bgroup \em et al.\egroup
  }{2015}]{Kulkarni2015}
Vivek Kulkarni, Rami Al-Rfou, Bryan Perozzi, and Steven Skiena.
\newblock Statistically significant detection of linguistic change.
\newblock In {\em 24th Intl. World Wide Web Conference}, pages 625--635. ACM,
  2015.

\bibitem[\protect\citeauthoryear{Leskovec and Krevl}{2014}]{snapnets}
Jure Leskovec and Andrej Krevl.
\newblock {SNAP Datasets}: {Stanford} large network dataset collection, 2014.

\bibitem[\protect\citeauthoryear{Leskovec \bgroup \em et al.\egroup
  }{2007}]{Leskovec2007}
Jure Leskovec, Jon Kleinberg, and Christos Faloutsos.
\newblock Graph evolution: Densification and shrinking diameters.
\newblock {\em ACM Transactions on Knowledge Discovery from Data (TKDD)},
  1(1):2, 2007.

\bibitem[\protect\citeauthoryear{Maaten and Hinton}{2008}]{Maaten2008}
Laurens van~der Maaten and Geoffrey Hinton.
\newblock Visualizing high-dimensional data using t-sne.
\newblock {\em Journal of Machine Learning Research}, 9(Nov):2579--2605, 2008.

\bibitem[\protect\citeauthoryear{Niepert \bgroup \em et al.\egroup
  }{2016}]{niepert2016learning}
Mathias Niepert, Mohamed Ahmed, and Konstantin Kutzkov.
\newblock Learning convolutional neural networks for graphs.
\newblock In {\em International Conference on Machine Learning}, pages
  2014--2023, 2016.

\bibitem[\protect\citeauthoryear{Ou \bgroup \em et al.\egroup }{2016}]{Ou2016}
Mingdong Ou, Peng Cui, Jian Pei, Ziwei Zhang, and Wenwu Zhu.
\newblock Asymmetric transitivity preserving graph embedding.
\newblock In {\em Proc. 22nd Intl. Conf. on Knowledge Discovery and Data
  Mining}, pages 1105--1114, 2016.

\bibitem[\protect\citeauthoryear{Park \bgroup \em et al.\egroup
  }{2009}]{Park2009}
Youngser Park, C~Priebe, D~Marchette, and Abdou Youssef.
\newblock Anomaly detection using scan statistics on time series hypergraphs.
\newblock In {\em Link Analysis, Counterterrorism and Security (LACTS)
  Conference}, page~9, 2009.

\bibitem[\protect\citeauthoryear{Perozzi \bgroup \em et al.\egroup
  }{2014}]{Perozzi2014}
Bryan Perozzi, Rami Al-Rfou, and Steven Skiena.
\newblock Deepwalk: Online learning of social representations.
\newblock In {\em Proc. 20th Intl. Conf. on Knowledge Discovery and Data
  Mining}, pages 701--710, 2014.

\bibitem[\protect\citeauthoryear{Roweis and Saul}{2000}]{Roweis2000}
Sam~T Roweis and Lawrence~K Saul.
\newblock Nonlinear dimensionality reduction by locally linear embedding.
\newblock {\em science}, 290(5500):2323--2326, 2000.

\bibitem[\protect\citeauthoryear{Sun \bgroup \em et al.\egroup
  }{2007}]{Sun2007}
Jimeng Sun, Christos Faloutsos, Spiros Papadimitriou, and Philip~S Yu.
\newblock Graphscope: parameter-free mining of large time-evolving graphs.
\newblock In {\em Proc. 13th Intl. Conf. on Knowledge Discovery and Data
  Mining}, pages 687--696, 2007.

\bibitem[\protect\citeauthoryear{Sutskever \bgroup \em et al.\egroup
  }{2013}]{Sutskever2013}
Ilya Sutskever, James Martens, George~E Dahl, and Geoffrey~E Hinton.
\newblock On the importance of initialization and momentum in deep learning.
\newblock In {\em Proc. 30th Intl. Conf. on Machine Learning}, pages
  1139--1147, 2013.

\bibitem[\protect\citeauthoryear{Tang \bgroup \em et al.\egroup
  }{2015}]{Tang2015}
Jian Tang, Meng Qu, Mingzhe Wang, Ming Zhang, Jun Yan, and Qiaozhu Mei.
\newblock Line: Large-scale information network embedding.
\newblock In {\em 24th Intl. World Wide Web Conference}, pages 1067--1077,
  2015.

\bibitem[\protect\citeauthoryear{Tenenbaum \bgroup \em et al.\egroup
  }{2000}]{Tenenbaum2000}
Joshua~B Tenenbaum, Vin De~Silva, and John~C Langford.
\newblock A global geometric framework for nonlinear dimensionality reduction.
\newblock {\em science}, 290(5500):2319--2323, 2000.

\bibitem[\protect\citeauthoryear{Wang and Wong}{1987}]{Yuchung1987}
Yuchung~J Wang and George~Y Wong.
\newblock Stochastic blockmodels for directed graphs.
\newblock {\em Journal of the American Statistical Association}, 82(397):8--19,
  1987.

\bibitem[\protect\citeauthoryear{Wang \bgroup \em et al.\egroup
  }{2016}]{Wang2016}
Daixin Wang, Peng Cui, and Wenwu Zhu.
\newblock Structural deep network embedding.
\newblock In {\em Proc. 22nd Intl. Conf. on Knowledge Discovery and Data
  Mining}, pages 1225--1234, 2016.

\bibitem[\protect\citeauthoryear{Zhu \bgroup \em et al.\egroup
  }{2016}]{Zhu2016}
Linhong Zhu, Dong Guo, Junming Yin, Greg Ver~Steeg, and Aram Galstyan.
\newblock Scalable temporal latent space inference for link prediction in
  dynamic social networks.
\newblock {\em IEEE Transactions on Knowledge and Data Engineering},
  28(10):2765--2777, 2016.

\end{thebibliography}

\end{document}